\documentclass[aps,pra,floatfix,nofootinbib,tightenlines,twocolumn,amsmath,amssymb]{revtex4}
\pdfoutput=1

\usepackage{amsmath,bbm,graphicx}
\usepackage[letterpaper=true,pdftex,pdfpagelabels=true,bookmarks = true,colorlinks = true, citecolor=blue, urlcolor=cyan]{hyperref}

\newcommand{\bra}[1]{\ensuremath{\left\langle #1\right\vert}}
\newcommand{\ket}[1]{\ensuremath{\left\vert #1\right\rangle}}
\newcommand{\expval}[1]{\ensuremath{\left\langle #1 \right\rangle}}
\newcommand{\hsp}[1]{\hspace{#1 em}}
\newcommand{\sqz}{\hsp{-0.1}}
\newcommand{\ketbra}[2]{\left\vert{#1}\right\rangle \sqz\sqz\sqz \left\langle{#2}\right\vert}
\newcommand{\braket}[2]{\left\langle{#1}\right\vert \sqz \sqz \left. {#2}\right\rangle}

\newcommand{\tr}{\text{Tr}}
\newcommand{\nbar}{\overline{n}}
\def\iden{\mathbbm{1}}

\graphicspath{{Figures/}}

\newtheorem{theorem}{Theorem}

\begin{document}
\title{Quantum benchmarking with realistic states of light}
\date{\today}

\author{N. Killoran${}^1$, M. Hosseini${}^2$, B. C. Buchler${}^2$, P. K. Lam${}^2$, and N. L\"{u}tkenhaus${}^1$}
\affiliation{${}^1$ Institute for Quantum Computing and Department of Physics \&
Astronomy, University of Waterloo, N2L 3G1 Waterloo, Canada}
\affiliation{${}^2$ Centre for Quantum Computation and Communication Technology, 
Department of Quantum Science, The Australian National University,
Canberra, ACT 0200, Australia.}

\begin{abstract}
The goal of quantum benchmarking is to certify that imperfect quantum communication devices (e.g., quantum channels, quantum memories, quantum key distribution systems) can still be used for meaningful quantum communication. However, the test states used in quantum benchmarking experiments may be imperfect as well. Many quantum benchmarks are only valid for states which match some ideal form, such as pure states or Gaussian states. We outline how to perform quantum benchmarking using arbitrary states of light. We demonstrate these results using real data taken from a continuous-variable quantum memory.
\end{abstract}

\maketitle


\section{Introduction}\label{sec:intro}

Quantum communication is poised to be the first real-world application of quantum information theory. Impressive efforts are underway to build advanced quantum devices, such as quantum channels \cite{wittmann10a}, quantum memories \cite{lvovsky09a}, and quantum key distribution systems \cite{lance05a}, which will make long-distance quantum communication a reality. But our control over the quantum world is still imperfect; it is difficult to create devices whose operation faithfully matches the ideal. Internal flaws and external influences can negatively impact a device's operation. Since quantum effects are very fragile, there is little tolerance for such deficiencies. Fortunately, it is still possible for imperfect devices to facilitate quantum communication.

The task of \emph{quantum benchmarking} is to certify a device to be within the ``quantum domain,'' the operational regime where an imperfect device is still useful for quantum communication. In practice, this is done by preparing some test states, probing the device in question with these states, and making measurements on the resulting output states. The corresponding theoretical task is to calculate the threshold (according to some figure(s) of merit) between a quantum device and the best comparible classical device. If we can conclude, based on the available information, that the device's performance beats the benchmark value, then the device is in the quantum domain. Of course, there is some interplay between the two sides, especially in the choice of test states.

A number of benchmarks have been developed in recent years, especially for continuous variable (CV) encodings of light \cite{braunstein00a,braunstein01a,grosshans01a,hammerer05a,namiki08a,namiki08b,adesso08a,owari08a,hetet08b,calsamiglia09a,guta10a, rigas06a, rigas06b, haseler08a, haseler09a, haseler10a, killoran10b, killoran11a}. Many of these benchmarks require test states which are pure and/or Gaussian. Such states are much more amenable to theoretical analysis than more general states of light. Now, if we recognize that our quantum communication devices are imperfect, we must also recognize that our state preparations may be imperfect as well (e.g., due to excess
noise or other flaws in the preparation procedure). This leads to discrepencies between the idealized benchmarking situation and the experimental realization. Using state tomography or some other process to obtain numerical descriptions of the prepared states, we may find that these states are not consistent with the ideal form (e.g., not pure or not Gaussian). In this case, is it valid to apply benchmarks which assume the ideal form? Furthermore, many of the above benchmarks require an infinite ensemble of test states. What conclusions can be made when using a finite ensemble (as we must always do in practice) if the benchmark requires an infinite ensemble? Addressing these practical concerns is an important part of quantum benchmarking.

In this work, we present a general benchmarking framework which is easily adaptable to different testing conditions. In particular, our framework allows for quantum benchmarking using arbitrary states, not just states with a special form. Previous progress has been made toward this goal using \emph{entanglement-based} benchmarks, but these schemes are still limited to special cases: either the test ensemble contains only pure states \cite{rigas06a,rigas06b,haseler08a,haseler10a}, or it consists of only two mixed states \cite{haseler09a} (in either case, the states are not required to be Gaussian). We extend these results to any finite number of mixed states, allowing us to derive quantum benchmarks for arbitrary finite ensembles of test states. There is some freedom allowed by this extension, and we outline how to strengthen benchmarks to give the largest quantum domain, i.e., the best chance of certifying imperfect quantum devices. We then apply our benchmarking criterion to an actual continuous-variable memory.

The remainder of this paper is laid out as follows. In Sec. \ref{sec:benchmarking} we outline the necessary theory. We show how to generalize previous results to an arbitrary number of test states and how to optimize this generalization. As an application, we consider phase-symmetric test ensembles, and derive a simple standard form for the bipartite matrices used to represent such ensembles in entanglement-based benchmarking. In Sec. \ref{sec:qmemtest}, we test our general benchmarking method using data from a real continuous-variable memory. Finally, in Sec. \ref{sec:conclusion}, we make some concluding observations.

\section{Entanglement-based quantum benchmarking}\label{sec:benchmarking}
\subsection{Benchmarking background}\label{ssec:benchbackground}

We picture any potential quantum communication device as a channel which takes quantum states as input and gives quantum states as output. The distinction between quantum and classical communication devices is that classical devices are equivalent to \emph{measure and prepare} (MP) channels. In these channels, the input state is measured, the measurement outcome is communicated or stored in a solely classical manner, and an output state is prepared based on this classical data. Any device that is equivalent to a MP channel is not a true quantum communication device, since communication through MP channels is equivalent to classical communication. It is often convenient to picture benchmarking in an adversarial way. In this scenario, we allow the possibility that a quantum device could have been replaced by a MP channel trying to pass our quantum benchmarking tests. Only by conclusively ruling out MP operation can we certify a quantum device as genuine. Devices which can be discriminated from classical channels are said to be in the quantum domain.

In practice, we probe the device with a finite number of different test states $\{\rho_k \}_{k=0}^{M-1}$ and make measurements on the resulting output states. From the measurement results and knowledge of the test ensemble, we must determine whether the device is in the quantum domain or whether the same operation could be performed by a classical channel. With enough test states and measurements, we can characterize a device exactly, i.e., perform process tomography, allowing a device's performance to be tested against any theoretical benchmark \cite{lobino08a,lobino09a}. However, process tomography can be resource intensive, especially for CV systems. Our goal will be to certify quantum devices with minimal requirements. If a given benchmarking test fails, it may still be possible to certify a quantum device by incorporating more test states and/or more measurements. 

There are a number of approaches for determining whether a device is in the quantum domain. Typically, some specific operational or theoretical quantities are chosen and the corresponding quantum domain is mapped out by determining the limits of classical devices. One such example is the state-independent $T$-$V$ benchmark based on signal-transfer and conditional-variance \cite{hetet08b, hosseini11a}, parameters which capture how well a signal is preserved and how much extra noise is introduced by the device. Another much-studied figure of merit is the average fidelity between the test states and the corresponding output states after the device action (represented by the map $\Lambda$):
\begin{equation}
	\overline{F} = \sum_{k=0}^{M-1}p_k F(\rho_k,\Lambda[\rho_k]),
\end{equation}
where $p_k$ are some fixed weights. If the test states are not perfectly distinguishable, a MP channel cannot achieve perfect fidelity. Indeed, for a given test ensemble, there is some maximum average fidelity $\overline{F}_{MP}^\text{max}$ which is achievable by MP channels. A device must beat this benchmark value, $\overline{F}>\overline{F}_{MP}^\text{max}$, to be in the quantum domain. This approach requires us to compute the optimal value $\overline{F}_{MP}^\text{max}$ and to calculate or bound the experimentally-achieved value $\overline{F}$. Typically, finding the experimental fidelity requires tomography on the output states. As well, tomography on the prepared states is needed to ensure that they match the required form. 

A number of fidelity-based benchmarks have been derived for CV test states, including coherent states, squeezed states, and thermalized states \cite{braunstein00a,braunstein01a,hammerer05a,namiki08a,namiki08b,adesso08a,owari08a,calsamiglia09a}. Other benchmarks use the same approach but slightly different figures of merit \cite{owari08a, guta10a} or models for classical channels \cite{grosshans01a}. Many of these benchmarks require test ensembles with an infinite number of states, although fidelity benchmarks for any two pure states have also been calculated \cite{namiki08a,fuchs03a}. The classical threshold $\overline{F}_{MP}^\text{max}$ is often very difficult to determine, making the fidelity-based approach hard to generalize to arbitrary test ensembles. This limits the applicability of this method to those ensembles where a benchmark value has been found, and we must verify that the experimentally-prepared states match the ideal form. As well, without (costly) tomography on the output states, or extra assumptions, it can be difficult to properly determine the experimental fidelity value $\overline{F}$.

There is another approach to quantum benchmarking \cite{rigas06a, rigas06b, haseler08a, haseler09a, haseler10a, killoran10b, killoran11a} which avoids many of these problems. This approach is based on the following important property: MP channels always break entanglement. In fact, the set of MP channels is equivalent to the set of entanglement-breaking channels \cite{horodecki03a}. By showing that a device preserves entanglement, we can exclude it from the MP class. To accomplish this goal, a bipartite entangled state, built from the test states, is considered. For the moment, we will focus on pure test states $\rho_k=\ketbra{\psi_k}{\psi_k}$, and we will generalize to mixed states in the next subsection. We imagine the test states are coupled with an ancillary system, giving the following entangled state
\begin{equation}\label{eq:basicentstate}
	\ket{\Psi^\text{ent}}_{AA'}=\frac{1}{\sqrt{M}}\sum_{k=0}^{M-1}\ket{k}_A\ket{\psi_k}_{A'}.
\end{equation}
Here, Alice's states lie in an $M-$dimensional Hilbert space for which $\{\ket{k}_A\}_{k=0}^{M-1}$ forms an orthonormal basis. We let the device in question act on the $A'$ subsystem of this entangled state, mapping it to some output system $B$, while subsystem $A$ is kept isolated. The reduced state $\rho_A$ remains the same, independent of the channel: 
\begin{equation}\label{eq:rhoAdef}
	\bra{k}\rho_A\ket{l} = \frac{1}{M}\braket{\psi_l}{\psi_k}.
\end{equation}
Aside from a fixed prefactor, the density matrix $\rho_A$ is the Gram matrix (matrix of overlaps) for the pure test states $\{\ket{\psi_k}\}$. This structure will be important when we generalize to arbitrary states in the next subsection.

Performing the local projective measurements $\{\hat{\Pi}_A^k:=\ketbra{k}{k}_A\}_{k=0}^{M-1}$ on $\ket{\Psi^\text{ent}}_{AA'}$ effectively prepares the test states $\{\ket{\psi_k}_{A'}\}_{k=0}^{M-1}$ at random. After passing through the channel, the output states $\{\rho_k^\text{out}:=\Lambda(\ketbra{\psi_k}{\psi_k})\}_{k=0}^{M-1}$ (on the system $B$), are measured with some fixed set of measurements $\{\hat{O}^j_B\}$. Using only these measurement results and knowledge of the Gram matrix $\rho_A$, we aim to determine whether the state $\rho_{AB}^\text{out}=\left[\text{id}_A\otimes\Lambda\right] \ketbra{\Psi^\text{ent}}{\Psi^\text{ent}}_{AA'}$ is entangled. 

One method to certify entanglement is with a witnessing procedure \cite{rigas06a, rigas06b, haseler08a, haseler10a}. On the other hand, it can be useful not only to qualify entanglement, but also to quantify it. An alternative approach to entanglement-based benchmarking \cite{killoran10b, killoran11a} is therefore to bound the entanglement of the final state away from zero using some suitable entanglement measure $\mathcal{E}$, i.e., to show $\mathcal{E}(\rho_{AB}^\text{out})>0$. One advantage of the quantitative method is that it can be used to compare different devices or situations which fall in the quantum domain. When we demonstrate our benchmarking method in Sec. \ref{sec:qmemtest}, we will take the quantitative approach, using tools from \cite{killoran11a}.

The measurements on system $A$ are restricted to the projections $\{\Pi_A^k\}_{k=0}^{M-1}$. Performing more general measurements on system $A$ would prepare superpositions of the test states. But these superposition states can simply be included in the test ensemble, offering the same formalism with a slightly different entangled state for Eq. (\ref{eq:basicentstate}). This source replacement scheme allows us to work with the test states in practice and consider the entangled state $\ket{\Psi}_{AA'}$ as a virtual construct. The Gram matrix $\rho_A$ provides the theoretical link between the test states and the virtual entangled state.

Finally, the Gram matrix compactly contains all information about the pure test states which is needed for the given benchmarking procedure. A full description of the test ensemble is not strictly necessary. Of course, we require some method for determining the overlap information. State tomography is one way to obtain this, but there may be other less costly ways to constrain the Gram matrix. Similarly, we do not require tomography on the output states for entanglement-based benchmarking. Indeed, previous papers have demonstrated that useful quantum benchmarks can be found by measuring only two conjugate quadratures \cite{rigas06a, rigas06b, lorenz06a, haseler08a, haseler09a,haseler10a,wittmann10a, killoran10b, killoran11a}. Thus, entanglement-based benchmarking can be very practical from an experimental point of view. 

\subsection{Entanglement-based picture for arbitrary states}

How can the entangled state in Eq. (\ref{eq:basicentstate}) be extended to the case where the test states $\{\rho_k\}_{k=0}^{M-1}$ are mixed? Some progress was made in this direction in \cite{haseler09a}, restricting to the case of only two test states. Among several potential ways to generalize Eq. (\ref{eq:basicentstate}), it was found that using purifications of the test states was the most useful. In this framework, we imagine that instead of preparing the test states, we prepare purifications $\{\ket{\Gamma_k}_{A'A''}\}_{k=0}^{M-1}$, where $A''$ is some purifying system, i.e.,
\begin{equation}
	\tr_{A''}(\ketbra{\Gamma_k}{\Gamma_k}_{A'A''}) = \rho_k.
\end{equation}
From now on, $\ket{\Gamma_k}$ will always refer to a purification of $\rho_k$. 

To clarify, the experimental procedure (using the test states $\{\rho_k\}$) remains the same. Theoretically, the benchmarking procedure is analyzed as if the purifications $\{\ket{\Gamma_k}_{A'A''}\}$ were used instead, i.e., the device is allowed to act on the larger system $A'A''$. Measurements are still restricted to the $B$ subsystem, and the only information we retain about the purifications for the final benchmarking are the overlaps $\braket{\Gamma_k}{\Gamma_l}$. In this way, we give extra power to an adversarial MP device, since the purifications may be more easily distinguished than the test states. Although this modification may result in a slightly smaller quantum domain, any benchmarks based on the purifications are valid. Now, even though we give extra power to an adversarial device, we should still aim to make our quantum benchmarks as stringent as we can, i.e., to keep the quantum domain as large as possible. Ref. \cite{haseler09a} offers one method to limit the adversarial advantage provided by purifications, but this introduces additional numerical complexity, and we will not consider it in the present paper. However, there still remains some freedom in the actual choice of purifications, which we can use to strengthen our benchmarking scheme. 

For two mixed test states $\rho_0$ and $\rho_1$, the virtual entangled state takes the form
\begin{equation}\label{eq:twomixed_entstate}
	\ket{\Psi}_{AA'A''} = \frac{1}{\sqrt{2}}\displaystyle\Big(\ket{0}_A\ket{\Gamma_0}_{A'A''}+\ket{1}_A\ket{\Gamma_1}_{A'A''}\Big).
\end{equation}
Purifications are not unique, but we must fix some choice to proceed further. Since we only retain information about the overlap $\braket{\Gamma_0}{\Gamma_1}$, it is important to know the range of allowed values for this parameter. We can always find purifications which are orthogonal. On the other hand, from Uhlmann's theorem \cite{uhlmann76a,jozsa94a}, we have 
\begin{equation}\label{eq:fidelbound}
	\max_{\ket{\Gamma_0},\ket{\Gamma_1}}\left|\braket{\Gamma_0}{\Gamma_1}\right|^2=\left[\tr\sqrt{\sqrt{\rho_0}\rho_1\sqrt{\rho_0}}\right]^2=:F(\rho_0,\rho_1),
\end{equation}
where $F(\rho_0,\rho_1)$ is the fidelity between the two test states. Thus, the range of allowed overlaps is $|\braket{\Gamma_0}{\Gamma_1}|^2\in[0,F(\rho_0,\rho_1)]$. In \cite{haseler09a}, it was argued that the best benchmarks (i.e., those which give the best opportunity to certify devices in the quantum domain) come from purifications which saturate the Uhlmann fidelity bound. There was numerical support for this claim, but one can also appeal to the fact that the fidelity is a measure of (in)distinguishability. Accordingly, choosing purifications which are as indistinguishable as possible will make it harder for a MP channel to simulate a quantum channel reliably. We will make this intuition more precise in the following pages. Finally, we note that Uhlmann's theorem has also been suggested for fidelity-based benchmarking using two mixed states \cite{namiki08a}.

It is straightforward to generalize the entangled state in Eq. (\ref{eq:twomixed_entstate}) to accomodate $M>2$ mixed states. Specifically, we will use
\begin{equation}\label{eq:Mmixed_entstate}
	\ket{\Psi}_{AA'A''} = \frac{1}{\sqrt{M}}\sum_{k=0}^{M-1}\ket{k}_A\ket{\Gamma_k}_{A'A''}.
\end{equation}
The reduced density matrix $\rho_{A}=\tr_{A'A''}\ketbra{\Psi}{\Psi}_{AA'A''}$ is now (aside from the prefactor) the Gram matrix of the purifications $\{\ket{\Gamma_k}\}_{k=0}^{M-1}$:
\begin{equation}\label{eq:rhoApurifs}
	\bra{k}\rho_A\ket{l}=\frac{1}{M}\braket{\Gamma_l}{\Gamma_k}.
\end{equation}
It is \emph{not} straightforward how to choose purifications in this case. The fidelity is a useful measure for two states, but there is no clear extension of the concept for multiple states. Indeed, the set of allowed values for the overlaps $\braket{\Gamma_k}{\Gamma_l}$ has a more subtle structure than just a product of the ranges allowed by Eq. (\ref{eq:fidelbound}). It may not be possible to choose purifications which achieve the fidelity bound for all pairs. In order to determine the best choice of purifications, we will examine the structure of entanglement-based benchmarking at a deeper level.

\subsection{Optimizing the purifications}\label{ssec:optimizepurifs}

How can we choose purifications to optimize the strength of entanglement-based benchmarks based on multiple mixed test states? To begin answering this, we make the following observation. 

\noindent\emph{Proposition 1. } Let $\{\ket{\Gamma_k}\}_{k=0}^{M-1}$ and $\{\ket{\Delta_k}\}_{k=0}^{M-1}$ be two different sets of purifications of the test states $\{\rho_k\}_{k=0}^{M-1}$. Assume that there is a completely positive, trace-preserving (CPTP) map which transforms one set of purifications to the other, i.e., there exists $\Omega$ such that
\begin{equation}\label{eq:ensemble_transf}
	\Omega\big[\ketbra{\Gamma_k}{\Gamma_k}\big]=\ketbra{\Delta_k}{\Delta_k}
\end{equation}
for all $k$. If we cannot conclude that a device is in the quantum domain with information based on the $\{\ket{\Delta_k}\}$, then we cannot conclude that the same device is in the quantum domain using information about the $\{\ket{\Gamma_k}\}$.

\noindent\emph{Proof. } Represent the device under investigation by the map $\Lambda$. This map takes in states on the joint $A'A''$ system and outputs states on the $B$ system. We proceed in two steps. First, we assume that we can perform tomographically complete measurements on the output states. Later we will relax this to incomplete measurements. In either case, we work under the constraint that, for all $k$,
\begin{equation}\label{eq:samechannelconst}
	\Lambda\big[\ketbra{\Delta_k}{\Delta_k}\big]=\Lambda\big[\ketbra{\Gamma_k}{\Gamma_k}\big]=\rho_k^\text{out},
\end{equation}
where $\rho_k^\text{out}$ is the output state corresponding to the test state $\rho_k$. This constraint is experimentally enforced, since the output states are independent of the purification we use to theoretically describe the protocol. Consider the case where we describe the test states using the purifications $\{\ket{\Delta_k}\}$. A benchmarking protocol is unsuccessful when we cannot discriminate between the device and a MP channel, i.e., when there is a MP channel which gives the same output states as the device. Assume then that there exists an MP channel $\tilde{\Lambda}_{MP}$ such that
\begin{equation}\label{eq:MPchannelaction}
	\tilde{\Lambda}_{MP}\big[\ketbra{\Delta_k}{\Delta_k}\big]=\rho_k^\text{out}
\end{equation}
for all $k$. Now consider the alternate situation where we use the purifications $\{\ket{\Gamma_k}\}$, and there exists a CPTP map $\Omega$ as in Eq. (\ref{eq:ensemble_transf}). We define another channel $\Sigma_{MP}$ by
\begin{equation}\label{eq:definesigma}
	\Sigma_{MP}:=\tilde{\Lambda}_{MP}\circ\Omega.
\end{equation}
The concatenation of any channel with a MP channel is as a MP channel, so $\Sigma_{MP}$ is in the MP class. From Eqs. (\ref{eq:ensemble_transf}-\ref{eq:MPchannelaction}), we must have
\begin{equation}
	\Sigma_{MP}\big[\ketbra{\Gamma_k}{\Gamma_k}\big]=\rho_k^\text{out}
\end{equation}
for all $k$. Therefore, there exists a MP channel which gives the observed output states when using the purifications $\{\ket{\Gamma_k}\}$.

Even when we do not have enough measurements for complete tomography, we can follow simular arguments. Instead of having one output state $\rho_k^\text{out}$ for every $k$, we have a set of states $\mathcal{C}[\rho_k^\text{out}]$, all of which have the same expectation values with respect to the employed measurement operators $\{\hat{O}_B^j\}$:
\begin{equation}
	\mathcal{C}[\rho_k^\text{out}] :=\big\lbrace\tau\big\vert\tr(\tau\hat{O}_B^j)=\langle\hat{O}_B^j\rangle_\text{meas}~\forall~j\big\rbrace.
\end{equation}
In this case, a benchmarking protocol is unsuccessful when there is a MP channel $\tilde{\Lambda}_{MP}$ such that
\begin{equation}\label{eq:MPunsucessful}
	\tilde{\Lambda}_{MP}\big[\ketbra{\Delta_k}{\Delta_k}\big]\in\mathcal{C}[\rho_k^\text{out}]
\end{equation}
for all $k$. Using this channel, we again define a MP channel $\Sigma_{MP}$ as in Eq. (\ref{eq:definesigma}). By Eqs. (\ref{eq:ensemble_transf}) and (\ref{eq:MPunsucessful}), we conclude
\begin{equation}
	\Sigma_{MP}\big[\ketbra{\Gamma_k}{\Gamma_k}\big]\in\mathcal{C}[\rho_k^\text{out}].
\end{equation}
In either case, if benchmarking is unsuccessful for the $\{\ket{\Delta_k}\}$, it cannot be successful for the $\{\ket{\Gamma_k}\}$.
\hfill$\square$

This proposition has some important consequences. For one, it tells us that in the source-replacement scheme we are using, the quantity of entanglement does not have a direct significance \cite{footnote1}. Instead, the focus should be to find those purifications which cannot be collectively transformed, via CPTP maps, to any other valid set. Benchmarking schemes built with such purifications provide the hardest challenge for an adversarial MP device attempting to mimic a true quantum device. Indeed, such limiting purifications necessarily lead to a larger quantum domain than any other comparable choice.  Because we only use the overlaps for benchmarking, we can optimize the CPTP map condition on the level of Gram matrices. In the rest of this paper, we will use the terms `purifications' and `Gram matrix (of the purifications)' somewhat interchangeably.

The following theorem will help us translate the CPTP map condition to a more amenable form.

\begin{theorem}[\cite{uhlmann85a, chefles00a, chefles04a}]
Let $\{\ket{\gamma_k}\}_{k=0}^{M-1}$ and $\{\ket{\delta_k}\}_{k=0}^{M-1}$ be two sets of pure states (they do not need to be purifications). Let $G$ and $D$ be the corresponding Gram matrices, with elements $G_{ij}:=\braket{\gamma_j}{\gamma_i}$ and $D_{ij}:=\braket{\delta_j}{\delta_i}$. There exists a CPTP map $\Omega$ taking the former states to the latter if and only if the Gram matrices are related by
\begin{equation}\label{eq:grammatrixcondition}
	G = P\circ D,
\end{equation}
where $\circ$ denotes the Hadamard (or Schur or entrywise) product. The matrix $P$ satisfies $P\geq 0$ and its diagonal elements are given by $\mathrm{diag}(P)=\{1,1,1,\dots\}$.
\end{theorem}

To avoid potential confusion, we point out that while the CPTP map $\Omega$ takes the $\{\ket{\gamma_k}\}$ to the $\{\ket{\delta_k}\}$, the Gram matrix condition, Eq. (\ref{eq:grammatrixcondition}), has the opposite sense, i.e., $G$ is obtained by doing a particular operation on $D$. Since some of the matrix elements of $P$ could be zero, we cannot invert the equation to give $D$ as a function of $G$, so we leave the relation in this form. We also point out a corollary to this theorem: all compatible purifications of the test states $\{\rho_k\}$ can be prepared by applying a CPTP map to a set of purifications which are orthogonal. Therefore, orthogonal purifications can be seen as generators for the rest of the set of purifications. Of course, we are interested in the other end of this generation, i.e., the limiting sets of purifications.

Given some ensemble of test states, it may be quite difficult to determine the best purifications analytically, especially for arbitrary test states. Alternatively, we can attempt to find limiting purifications by maximizing some objective function $f$, defined on ensembles of pure states (or on the corresponding Gram matrix), which preserves the order structure induced by CPTP maps. In other words, if there exists a CPTP map $\Omega$ taking the pure states $\{\ket{\gamma_k}\}_{k=0}^{M-1}$ to the pure states $\{\ket{\delta_k}\}_{k=0}^{M-1}$, the desired function must satisfy
\begin{equation}
	f(\{\ket{\gamma_k}\}) \leq f(\{\ket{\delta_k}\}).
\end{equation}
When this property holds, the purifications which are limiting in the sense of CPTP maps will maximize the objective function $f$. Before discussing candidates for the objective function, we point out one caveat. Namely, the order structure imposed on the purifications by CPTP maps is not a total order. Accordingly, purifications which are not linked by a CPTP map are not comparable. Nevertheless, we can design quantum benchmarking schemes with any feasible set of purifications, and this heuristic provides a way to obtain strong candidates.

A good candidate for the objective function is the \emph{Gram matrix purity} $\mathcal{P}$ (recall from Eq. (\ref{eq:rhoApurifs}) that $\rho_A$ is essentially the Gram matrix of the chosen purifications):
\begin{align}\label{eq:puritydef}
	\mathcal{P} = \tr(\rho_A^2)=&\sum_{k=0}^{M-1}\sum_{l=0}^{M-1}\left|\left[\rho_A\right]_{kl}\right|^2\nonumber\\
	=&\frac{1}{M^2}\sum_{k=0}^{M-1}\sum_{l=0}^{M-1}\left|\braket{\Gamma_k}{\Gamma_l}\right|^2.
\end{align}
From the above theorem, when there is a CPTP map taking $\{\ket{\Gamma_k}\}\rightarrow \{\ket{\Delta_k}\}$, we must have
\begin{equation}
	\braket{\Gamma_k}{\Gamma_l}=P_{kl}\braket{\Delta_k}{\Delta_l},
\end{equation}
with $|P_{kl}|\leq 1$. Therefore the Gram matrix purity is monotonic with respect to CPTP maps, as required. 

The Gram matrix purity also has links to the distinguishability of the test states. For example, when working with only two test states $\rho_0$ and $\rho_1$, the maximal value of $\mathcal{P}$ is
\begin{align}
	\max\mathcal{P} = &\frac{1}{4}\left(2+2\max_{\ket{\Gamma_0},\ket{\Gamma_1}}\left|\braket{\Gamma_0}{\Gamma_1}\right|^2\right)\\
	= &\frac{1}{2}\left(1+F(\rho_0,\rho_1)\right).
\end{align}
Aside from a fixed affine transformation, the Gram matrix purity is the fidelity between the test states. For $M>2$, the Gram matrix purity defines a kind of averaged multi-state analog of the fidelity. Finally, consider the state 
\begin{equation}
	\rho_{A'A''} = \frac{1}{M}\sum_{k=0}^{M-1}\ketbra{\Gamma_k}{\Gamma_k}_{A'A''}.
\end{equation}
Since a device has no information about which test state was prepared, this is the effective state input to the device in the purification picture. But since the virtual entangled state in Eq. (\ref{eq:Mmixed_entstate}) is pure, the spectra of the reduced states are equal, i.e., $\text{spec}(\rho_A) = \text{spec}(\rho_{A'A''})$. Therefore, when $\rho_A$ is of high purity, then so is the average test state $\rho_{A'A''}$. Intuitively, when the average state $\rho_{A'A''}$ is close to a pure state, its constituant test states $\{\ketbra{\Gamma_k}{\Gamma_k}\}$ will be difficult to distinguish.

One drawback of the Gram matrix purity is that it will be difficult to compute the maximum for $M>2$, where we do not currently have an analytic formula like Eq. (\ref{eq:fidelbound}). Alternatively, since the objective function in Eq. (\ref{eq:puritydef}) is convex and the set of compatible Gram matrices is convex (see Appendix \ref{app:grammatconv}), we could maximize this function numerically using the methods of convex optimization. Although this presents one path towards our goal, we will not pursue it here.

Instead, the Gram matrix purity helps us motivate a slightly different objective function which is easier to optimize numerically. To write this alternate function, we abbreviate the overlaps by 
\begin{equation}
	Z_{kl} := \braket{\Gamma_k}{\Gamma_l}
\end{equation}
and decompose them into real and imaginary parts:
\begin{equation}
	Z_{kl} = X_{kl}+iY_{kl}.
\end{equation}
The alternate objective function is given by
\begin{equation}\label{eq:modobjfun}
	h:= \sum_{k=l+1}^{M-1}\sum_{l=0}^{M-1}\left(X_{kl}+Y_{kl}\right).
\end{equation}

Aside from a prefactor and duplicated terms, the main difference between $h$ and the Gram matrix purity $\mathcal{P}$ is that the squared modulus of the overlaps is replaced by a sum of the real and imaginary parts. The advantage of this objective function is that it is linear in the parameters $X_{kl}$ and $Y_{kl}$, allowing us to numerically optimize it using a semidefinite program, which can be done efficiently. In cases where the the imaginary parts $Y_{kl}$ vanish, then $X_{kl}+Y_{kl}$ is a lower bound to the modulus $|Z_{kl}|$. As well, when the objective function $h$ is large, the Gram matrix purity will also be large, meaning we are close to a limiting Gram matrix. The main drawback of this objective function is that it is not always monotonic. Nevertheless, we remind the reader that \emph{any} compatible Gram matrix, including one found by an optimization process over $h$, is valid for benchmarking purposes. Indeed, we will see in Sec. \ref{sec:qmemtest} that, despite being non-monotonic, this function leads to a Gram matrix with near-optimal purity.

Before moving on, we pause to summarize the main results of this section. Given some arbitrary ensemble of test states $\{\rho_k\}_{k=0}^{M-1}$, we consider a virtual entangled state as in Eq. (\ref{eq:Mmixed_entstate}), where the states $\{\ket{\Gamma_k}_{A'A''}\}_{k=0}^{M-1}$ are purifications of the test states. Although the test states are used in practice, we benchmark a device by assuming that the device has access to the purifications. This framework provides more power to an adversarial device, but there is also some flexibility in the choice of purifications, which can be used to restrict the adversarial advantage. We showed that purifications which are extremal in the sense of CPTP maps are better for benchmarking than all other comparable choices, so we should aim to build benchmarks using such purifications. We suggest finding good candidate purifications by optimizing some appropriate function which is monotonic with respect to CPTP maps. We propose the Gram matrix purity $\mathcal{P}$ as a suitable candidate for this objective function, although it may be difficult to compute. We also propose a numerically simpler, though non-monotonic, objective function which may be used in place of the Gram matrix purity. After the purifications have been chosen, the benchmarking proceeds as in the pure state case (see Sec. \ref{ssec:benchbackground}), requiring only the associated Gram matrix.

\subsection{Benchmarking with phase-symmetric ensembles}

One useful application for the above results is in testing devices which are phase covariant. A device (represented by $\Lambda$) is phase covariant when it commutes with unitary rotations of phase space, i.e.,
\begin{equation}
	\Lambda[U_{\phi}\sigma U_{\phi}^\dagger] = U_{\phi}\Lambda[\sigma ]U_{\phi}^\dagger~\forall~\sigma.
\end{equation}
The rotation unitary is given by $U_{\phi} = \exp(-i\phi\hat{n})$, with $\hat{n}$ the standard number operator, so $U_{\phi}^\dagger=U_{-\phi}$. We note that any channel can be made phase covariant by phase-randomization,
\begin{equation}\label{eq:phaserandomize}
	\Lambda_{C}[~\cdot~] = \frac{1}{2\pi}\int_0^{2\pi}U_{\phi}\Lambda[U_{-\phi}(~\cdot~)U_{-\phi}^\dagger]U_{\phi}^\dagger d\phi.
\end{equation}
One way to accomplish this phase randomization is to use a drifting optical phase \cite{haseler10a}, a situation which is common in many continuous-variable setups. If the phase-randomized version of a channel passes a quantum benchmark, then the channel itself must be in the quantum domain as well. On the other hand, since $\Lambda_{C}$ involves a concatenation of channels, a phase-randomized channel may perform weaker against our benchmarks than the original channel. 

Phase-covariant channels offer a number of advantages. On the experimental side, they can be benchmarked using only one physical test state. The effects of a phase-covariant channel on many other test states can be inferred by symmetry. On the theoretical side, phase covariance can lead to great numerical simplifications. If the test states are unrelated, the number of free parameters scales quadratically with the number of states $M$. When phase symmetry conditions are imposed, we will show that the number of free parameters can be made to scale only linearly with $M$. This gives us more computational room to push up the number of states $M$ in the test ensemble, leading to stronger benchmarks. 

To benchmark a phase-covariant channel, we consider test states from the rotationally-symmetric ensemble 
\begin{equation}\label{eq:rhorotations}
	\{\rho_k | \rho_k = U_{\theta}^k\rho_0 U_{\theta}^{\dagger k}\}_{k=0}^{M-1},
\end{equation}
where $\theta=\tfrac{2\pi}{M}$ (the angle $\theta$ will always be defined this way). For phase-covariant devices, we have
\begin{equation}\label{eq:phasecov}
	\Lambda[\rho_k] = U_{\theta}^k\Lambda[\rho_0] U_{\theta}^{\dagger k},
\end{equation}
so we can infer the channel's effect on $\rho_k$ by suitable rotations of its action on one real test state $\rho_0$. In this way we can generate multi-state benchmark data efficiently from any seed state. We will refer to the situation where both the test states are phase symmetric and the tested device is phase covariant as \emph{phase-symmetric benchmarking}.

We will now present two standard form results relevant to phase-symmetric benchmarking. Given a particular device (mapping from system $A'A''$ to $B$), the output state $\rho_{AB}^\text{out}$ is defined by
\begin{equation}\label{eq:rhoABout}
	\rho_{AB}^\text{out}:=\left[\text{id}_A\otimes\Lambda\right] \ketbra{\Psi^\text{ent}}{\Psi^\text{ent}}_{AA'A''},
\end{equation}
where $\ket{\Psi^\text{ent}}_{AA'A''}$ is the effective entangled state in Eq. (\ref{eq:Mmixed_entstate}). We must show that this reduced state is entangled in order to certify the given device. Without loss of generality, we can decompose this state as
\begin{equation}\label{eq:outputstate}
	\rho_{AB}^\text{out} = \frac{1}{M}\sum_{k=0}^{M-1}\sum_{l=0}^{M-1}\ketbra{k}{l}_A\otimes\rho_{kl}^\text{out},
\end{equation}
where $\rho_{kl}^\text{out}$ are square matrices on the $B$ subsystem (the diagonal blocks $\rho_{kk}^\text{out}=\rho_{k}^\text{out}$ are the output states). In a phase-symmetric benchmarking situation, the following relation can be imposed:
\begin{equation}\label{eq:rotationsymm}
	U_\theta \rho_{kl} U_\theta^\dagger = \rho_{k+1,l+1}~(\text{mod } M)~\forall~k,l.
\end{equation}
States with this symmetry can be brought into a simple standard form: 

\begin{theorem}[Standard form]\label{thm:stform}
Let
\begin{equation}
	\tau_{AB} = \frac{1}{M}\sum_{k=0}^{M-1}\sum_{l=0}^{M-1}\ketbra{k}{l}_A\otimes\tau_{kl},
\end{equation}
be an arbitrary bipartite matrix. Let $\theta=\tfrac{2\pi}{M}$ and let $\omega_M=\exp(i\theta)$ denote the primitive $M$th root of unity. Assume that the following symmetry relation holds:
\begin{equation}\label{eq:taurotations}
	U_\theta \tau_{kl} U_\theta^\dagger = \tau_{k+1,l+1}~(\text{mod } M) ~\forall~k,l.
\end{equation}
Then $\tau_{AB}$ is unitarily equivalent to a block diagonal matrix 
\begin{equation}
	\mathcal{D}(\tau_{AB}) = \bigoplus_{k=0}^{M-1} E_k
\end{equation}
where 
\begin{equation}
	E_k = \frac{1}{M}\sum_{l=0}^{M-1}\omega_M^{k\cdot l}\tau_{kl}U_\theta^{l}.
\end{equation}
Moreover, 
\begin{equation}\label{eq:tau00decomp}
	\sum_{k=0}^{M-1}E_k=\tau_{00},
\end{equation}
and $\tau_{AB}$ is positive semidefinite if and only if all of the $E_k$ are positive semidefinite.

\end{theorem}
 
\noindent\emph{Proof.} See Appendix \ref{app:stformproof}.

In the next section, we will quantify the entanglement of the state $\rho_{AB}^\text{out}$ by computing lower bounds on the negativity \cite{zyczkowski98a,lee00a,vidal02a,plenio05a}, an entanglement measure defined by
\begin{equation}\label{eq:defineneg}
	\mathcal{N}(\tau_{AB}) = \frac{||\tau_{AB}^{T_A}||_1-\tr(\tau_{AB})}{2},
\end{equation}
where $T_A$ denotes partial transposition and $||\cdot||_1$ is the trace norm. For phase-symmetric benchmarking, we can greatly simplify this calculation.

\begin{theorem}[Trace norm in standard form]\label{thm:tracenormstform}
Let $\tau_{AB}$ be as in Theorem \ref{thm:stform}. Then the trace norm of the partially transposed state $\tau_{AB}^{T_A}$ reduces to the form
\begin{equation}\label{eq:negstform1}
	||\tau_{AB}^{T_A}||_1 = \sum_{k=0}^{M-1}||\tilde{E}_k||_1,
\end{equation}
where the $\{\tilde{E}_k\}_{k=0}^{M-1}$ are formed by rearranging the matrix elements of $\{E_k\}_{k=0}^{M-1}$ from the standard form of $\tau_{AB}$. Specifically, the matrix elements in the Fock basis are determined via
\begin{equation}\label{eq:negstform2}
	[\tilde{E}_k]_{jl}=[E_{j+l-k}]_{jl}~(\text{mod }M).
\end{equation}

\end{theorem}

\noindent\emph{Proof.} See Appendix \ref{app:tnormstformproof}.

These two standard form results allow us reduce the numerical complexity of phase-symmetric benchmarking. Although the constraints remain unchanged (the Gram matrix $\rho_A$ and measurements on the seed state $\rho_0$), the benchmarking state is more compactly encoded by using the $M$ square matrices $E_k$ instead of the $\frac{M(M+1)}{2}$ independent blocks $\tau_{kl}$. 

\section{Demonstration with real memory data}\label{sec:qmemtest}

We now demonstrate the above theoretical results using real data generated by a continuous-variable memory. This memory is based on a three-level lambda gradient echo scheme involving hyperfine spin states of warm rubidium atoms. Its fundamental operation has been outlined in detail elsewhere \cite{hosseini12a}. The data used in the present paper is based on the results reported in \cite{hosseini11a}. 
This memory is not expected to have any sensitivity to phase. Indeed, the equations governing the memory operation are phase-independent \cite{hosseini12a}. As well, there is no experimental indication of phase sensitivity in the memory, i.e., the observed loss and excess noise are uniform with respect to the angle in phase space. Furthermore, there is no cross-talk between consecutive runs of the experiment. A data set contains measurements from 100,000 pulses stored in, and then recalled from, the memory. Each pulse was separated in time by 0.1~ms which, given the decoherence rate of the atoms in the hot gas cell, means that each pulse was stored in a fresh ensemble of atoms. 

The experimental phase reference was allowed to drift freely between runs. In principle, this drift could be used to phase-randomize the device following the approach of \cite{haseler10a}, rendering the channel phase covariant. For the experimental data used in the present analysis, the observed drift was not fast enough to ensure the uniformly random distribution of phase angles required in Eq. (\ref{eq:phaserandomize}). In future experiments, uniformity of the phase drift could be imposed by modifying the repetition rate of the experiment (see Sec. \ref{ssec:dataacq} for more specific details).

However, given the very strong evidence that the memory is phase-insensitive and that consecutive runs are uncorrelated, it is reasonable to expect that the memory operation is intrinsically phase-covariant. Our goal here is to demonstrate the strength of our benchmarking approach under a variety of scenarios, by using data from a real memory. We therefore assume that running the experiment with a completely uniform phase drift would lead to the same data as obtained at the current repetition rate. In other words, the obtained experimental data is assumed to come from a phase-covariant memory. The phase-covariance assumption, which is independent from the benchmarking theory, allows us to easily generate a number of different test ensembles using data from a single seed state, i.e., we have many non-orthogonal test states $\{\rho_k | \rho_k = U_{\theta}^k\rho_0 U_{\theta}^{\dagger k}\}_{k=0}^{M-1}$. To fully certify the memory with no assumptions (i.e., where we treat the memory as a black box) would require a more uniform phase drift than we obtained. Nevertheless, the current scenario allows for a useful demonstratation of our benchmarking tools.

\subsection{Experimental data acquisition}\label{ssec:dataacq}

During an experimental run, light is generated from a continuous-wave master laser and split into a coupling field (for activating the memory), a strong continuous-wave local oscillator, a weak signal pulse, and a memory-reference pulse. The memory-reference pulse is detuned from the active frequencies of the memory (while maintaining a well-defined phase relation with the signal pulse), so that it passes through the memory without absorption. The signal pulse is sent to the memory shortly after the reference pulse such that the phase drift between the two pulses can be neglected. Thus, we can obtain phase information for the input signal by determining the phase of the memory-reference pulse. 

Both the output signal and memory-reference pulses are interfered with the local oscillator (which has not passed through the memory) in a homodyne measurement scheme. Data was collected using a 12-bit resolution data acquisition system. The homodyne measurement angle is varied between runs by allowing the signal and local oscillator to freely drift. This phase angle is identified by fitting a sine wave to the memory-reference pulse. A quadrature value was assigned to each signal pulse after interference with the local oscillator by integrating the amplitude of the pulse over its duration. Note that any inherent frequency shift due to the memory was cancelled coherently by adjusting the external magnetic field applied to the memory before data collection. This process generates quadrature measurement data for the output state $\rho_{0}^\text{out}$. Similar data is obtained for the input state $\rho_{0}$ by measuring it after it passes through an inactive memory \cite{footnote2}. 

Over repeated runs, many data points are collected to form a raw data set covering many phase angles. To ensure complete randomisation of the input pulse, the drift resulting from thermal fluctuations needs to be much faster than the repetition rate of the experiment. In our experimental setup, the pulse's thermal drift occurred at a rate on the order of Hz, while the experimental data was taken at a kHz rate. Running the experiment at a slower repetition rate would allow the phase reference enough time to fully randomize, making the phase-randomization procedure of \cite{haseler10a} possible and guaranteeing phase-covariance. Of course, this would also affect the time needed to acquire the experimental data.


Tomographic reconstruction was performed using an iterative maximum-likelihood algorithm \cite{rehacek01a} using data from 100,000 pulses, giving density matrices for both the input and output states in the Fock basis. A constant memory-induced phase offset \cite{hetet08c} between the input and output signals was observed, and cancelled digitally during the tomography step. The tomographically-reconstructed density matrices were numerically truncated after the first 30 Fock states ($\ket{0},\dots,\ket{29}$). This cutoff is supported by the fact that both the input state and output state are essentially confined to the first 10 Fock levels, for which the cutoff error in either state is $\approx10^{-5}$. From the tomography, the input state was found to have mean photon number $\expval{\hat{n}_\text{in}}=0.67$ \cite{hosseini11a}. This value is consistent with the value obtained using direct quadrature measurement \cite{webb06a}. We do not explicitly categorize this state, but it is qualitatively a `coherent state with added noise'. This extra noise is primarily due to amplitude fluctuations and changes in mode-matching occuring during the testing process. The tomographically-reconstructed output state had a similar form, with mean photon number $\expval{\hat{n}_\text{out}}=0.57$.


As stated in Sec. \ref{ssec:benchbackground}, output state tomography is not required in quantum benchmarking; measurement of two conjugate quadratures may be sufficient. To compare the two approaches, we also consider benchmarks using measurements from only $\hat{x}$ and $\hat{p}$. To get the required expectation values, we collected the original quadrature/phase data into bins containing 500 data points and restricted our attention to two bins corresponding to phase angles of $0$ and $90$ degrees. Taking the mean and standard deviation of the data within these bins, we get estimates of $\expval{\hat{x}}$, $\expval{\hat{p}}$, $\text{Var}(\hat{x})$, and $\text{Var}(\hat{p})$, where $\text{Var}(\hat{z})=\expval{\hat{z}^2}-\expval{\hat{z}}^2$ for an operator $\hat{z}\in\{\hat{x},\hat{p}\}$. Error bars, based on the finite sample size of the bins, are calculated using standard error propagation techniques. The final quadrature moment values are listed in Table \ref{expvalues}. 

\begin{table}
\centering
 \begin{tabular}{| c | c | c |}
\hline
 Operator & First moment $\expval{\cdot}$ & Second moment $\expval{\cdot}^2$\\
 \hline
 $\hat{x}$ & 0.01 $\pm$ 0.03 & 0.57 $\pm$ 0.04\\
\hline
 $\hat{p}$ & -0.95 $\pm$ 0.03 & 1.41 $\pm$ 0.09\\
\hline
\end{tabular}
\caption{Experimentally-determined first and second moments for the conjugate quadratures $\hat{x}$ and $\hat{p}$. The vacuum state would have variances $\text{Var}(\hat{x})=\text{Var}(\hat{p})=\tfrac{1}{2}$ in these units.}\label{expvalues}
\end{table}

\subsection{Finding the Gram matrix}

The test ensemble is generated by rotations of the seed state $\rho_0$, as in Eq. (\ref{eq:rhorotations}), forming a ring in phase space. Our theoretical description of the seed state comes from the tomographic reconstruction with the finite cutoff in Fock space. We make no further assumptions about the seed state other than that the tomographic description is accurate. Purifications, and the associated Gram matrix, will be based on this numerical description. Due to the mixed nature of this state, previous entanglement-based benchmarking schemes would be limited to only two test states. We will see later that two test states are not sufficient in this case to reliably (i.e., within experimental error) give non-zero output entanglement, but adding more test states enables a successful benchmarking. 

In order to find candidate Gram matrices for the purifications, we convert the optimization to a semidefinite program. To this end, we consider the bipartite matrix 
\begin{align}
	\rho_{AA'}^\text{in} = & \tr_{A''}\ketbra{\Psi^\text{ent}}{\Psi^\text{ent}}_{AA'A''}\nonumber\\
	= & \frac{1}{M}\sum_{k=0}^{M-1}\sum_{l=0}^{M-1}\ketbra{k}{l}_A\otimes\rho_{kl}^\text{in}.
\end{align}
The diagonal blocks of $\rho_{AA'}^\text{in}$ are the test states, i.e., $\rho_{kk}^\text{in}=\rho_k$, and the off-diagonal blocks $\rho_{kl}^\text{in}$ consist of free parameters. The off-diagonal blocks are linked to the purifications through the relation
\begin{equation}
	\tr(\rho_{kl}^\text{in})=M\bra{k}\rho_A\ket{l}=\braket{\Gamma_l}{\Gamma_k}.
\end{equation}
We have traced out the purifying system $A''$ since no measurements are performed on that system, but we have kept subsystem $A'$ in order to later ascribe some quantity of entanglement to the input state. 

As stated in Sec. \ref{ssec:optimizepurifs}, instead of using the Gram matrix purity $\mathcal{P}$, we optimized the slightly different and numerically more simple objective function $h$ from Eq. (\ref{eq:modobjfun}). Although it might be more elegant to use the Gram matrix purity $\mathcal{P}$, nothing in the benchmarking procedure depends on that function. In fact, the Gram matrix found by optimizing $h$ over the given experimental data works quite well for our purposes. To see this, we can consider simple lower and upper bounds on the maximum value of $\mathcal{P}$ compatible with the given constraints.

\begin{figure}
	\includegraphics[width= 1.0 \columnwidth]{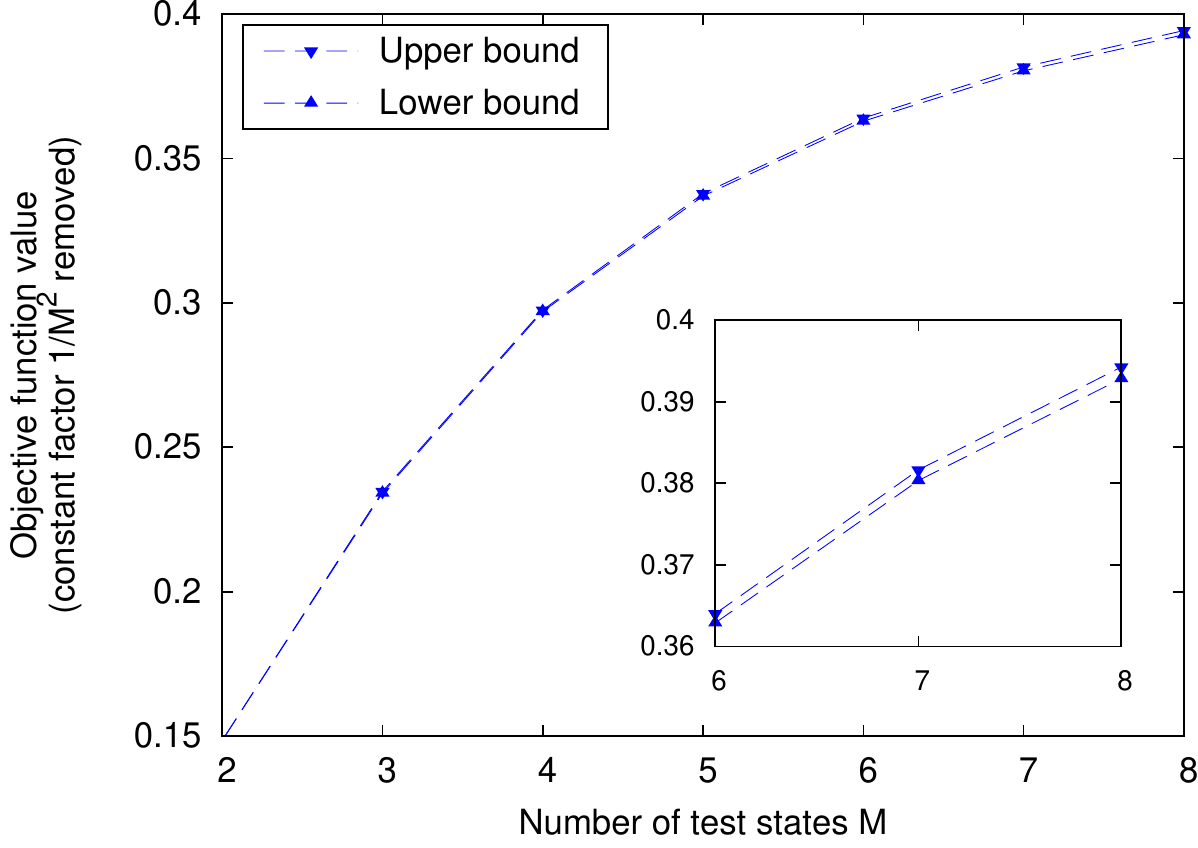}
	\caption{Upper and lower bounds for the maximum purity $\mathcal{P}$ of experimentally-compatible Gram matrices, corresponding to rotationally symmetric test ensembles of different cardinality. We have removed the factor $\tfrac{1}{M^2}$, which is common to both the upper and lower bounds, and independent of the purifications. The difference between the two bounds is on the order of $10^{-3}$ or less (inset), meaning that the obtained Gram matrix has purity which is near-optimal (represented by the lower curve). The estimated numerical precision of the optimization is $\approx 10^{-3}$.}
	\label{fig:objfuncompare}
\end{figure}

For one, the purity of the Gram matrix that maximizes $h$ (denoted $\arg\max h$) is less than the maximum possible Gram matrix purity over all purifications, i.e., 
\begin{equation}
	\left.\mathcal{P}\right|_{\arg \max h}\leq \max\mathcal{P}.
\end{equation}
On the other hand, we can easily find an upper bound on $\mathcal{P}$ by considering the pairwise fidelities:
\begin{equation}
	\max \mathcal{P} \leq \frac{1}{M^2}\sum_{k=0}^{M-1}\sum_{l=0}^{M-1}F(\rho_k,\rho_l).
\end{equation}
In Fig. \ref{fig:objfuncompare}, we plot the values of these lower and upper bounds for different numbers of rotationally symmetric test states (omitting the constant factors $\tfrac{1}{M^2}$). The computed bounds are very close to each other, differing only in the third decimal place, which is the same order as the estimated numerical precision of our optimization. We conclude that using the modified objective function in Eq. (\ref{eq:modobjfun}) is suitable for the situation we are considering, and that the obtained Gram matrix purity is close to what would be obtained by maximizing $\mathcal{P}$ itself.

\subsection{Computing the entanglement}

Having used the heuristic function $h$ to find a feasible $\rho_{AA'}^\text{in}$ which corresponds to a near-optimal Gram matrix, we now consider the entanglement of the corresponding output state $\rho_{AB}^\text{out}$. We use the negativity as our entanglement measure, which can be computed as a semidefinite program, namely
\begin{equation}
	\mathcal{N}(\tau_{AB})=	\displaystyle\min_{\tau_-\in \Upsilon} \tr(\tau_-),
\end{equation}
where the constraint set $\Upsilon$ is defined by
\begin{align}
 \Upsilon:=\{\tau_-|\tau_-\geq 0~\text{and}~\tau_{AB}^{T_A} + \tau_- \geq 0 \}.
\end{align}
For optimizations where $\tau_{AB}$ is not fully known (as is the situation here), constraints from experimental observations are also included in this set. For states written in standard form, the negativity can be computed on the level of the individual matrices $E_k$. 

Next, we used the entanglement quantification tools from \cite{killoran11a} to compute the lowest value of the negativity that could be compatible with the given data. In this approach, one rigorously finds lower bounds on the negativity of the output state $\rho_{AB}^\text{out}$ by truncating the $B$ subsystem of the density matrix at some finite Fock level $N$, which is large compared to the observed mean photon number $\nbar$. For this test, we imposed the cutoff at Fock level $N=15$, making the truncated matrices $16$-dimensional. 

The quantification procedure was applied to four measurement scenarios: 
\begin{enumerate}
\renewcommand{\labelenumi}{(\roman{enumi})}
 \item \textbf{Tomography}: the output density matrix is fully known numerically (up to the truncation level N); 
 \item \textbf{Quadratures from tomography}: only measurements on conjugate quadratures $\hat{x}$ and $\hat{p}$ for the output state are used, and these values are obtained from the tomographically-reconstructed density matrix; 
 \item \textbf{Quadratures from data}: same as (ii), but the quadrature values are obtained directly from the homodyne data;
 \item \textbf{Quadratures with error bars}: the same quadrature data is used as in (iii), but estimates of data error bars are included in the computation ($\sigma-3\sigma$ levels). 
\end{enumerate}

\begin{figure}[t]
	\includegraphics[width= 1.0 \columnwidth]{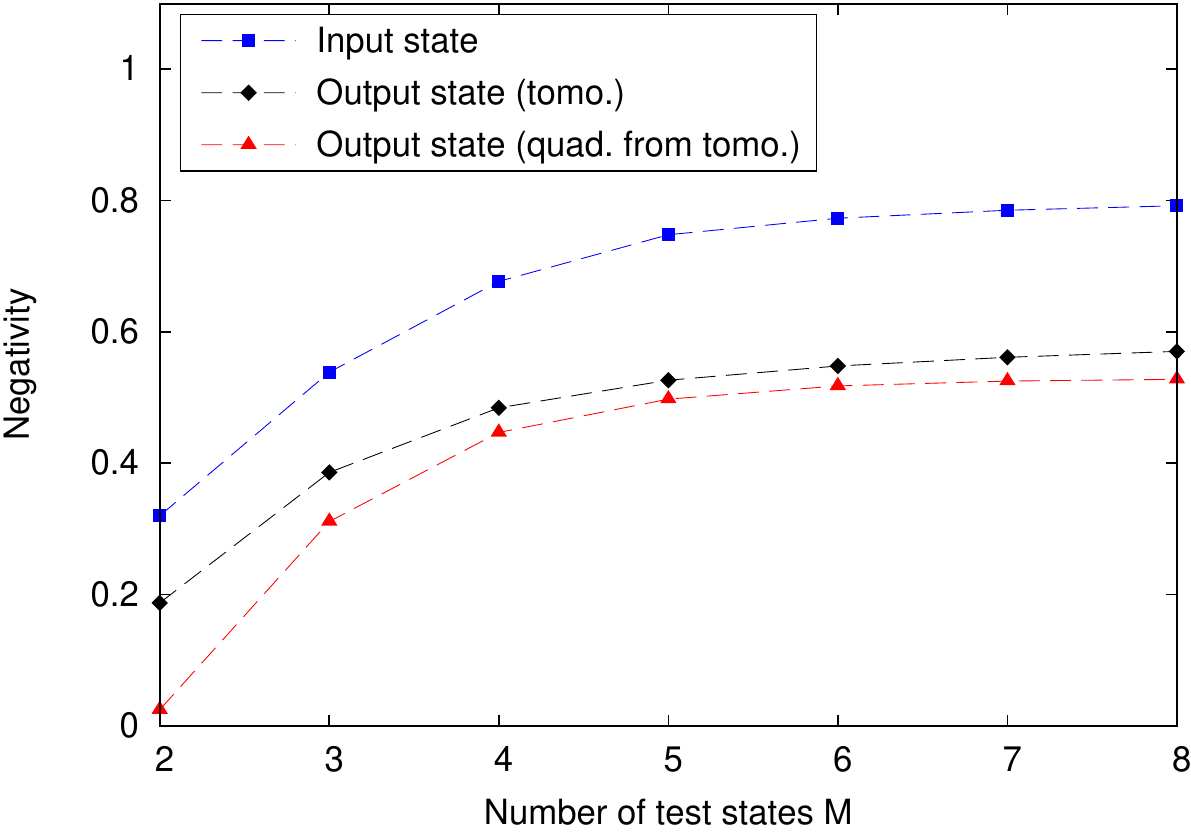}
	\caption[Benchmarking results using tomographically-reconstructed density matrices]{Benchmarking results based on rotationally-symmetric test ensembles using tomographically-reconstructed density matrices. The top curve is the negativity of the input state $\rho_{AA'}^\mathrm{in}$ found by optimizing the heuristic function $h$. The other curves are lower bounds to the negativity of $\rho_{AB}^\mathrm{out}$ based on (in descending order): (i) tomographic reconstruction and (ii) quadrature values obtained from the tomographic reconstruction. A benchmarking procedure is successful when the output state has negativity larger than zero.}
	\label{fig:negtomo}
\end{figure}

\begin{figure}[t]
	\includegraphics[width= 1.0 \columnwidth]{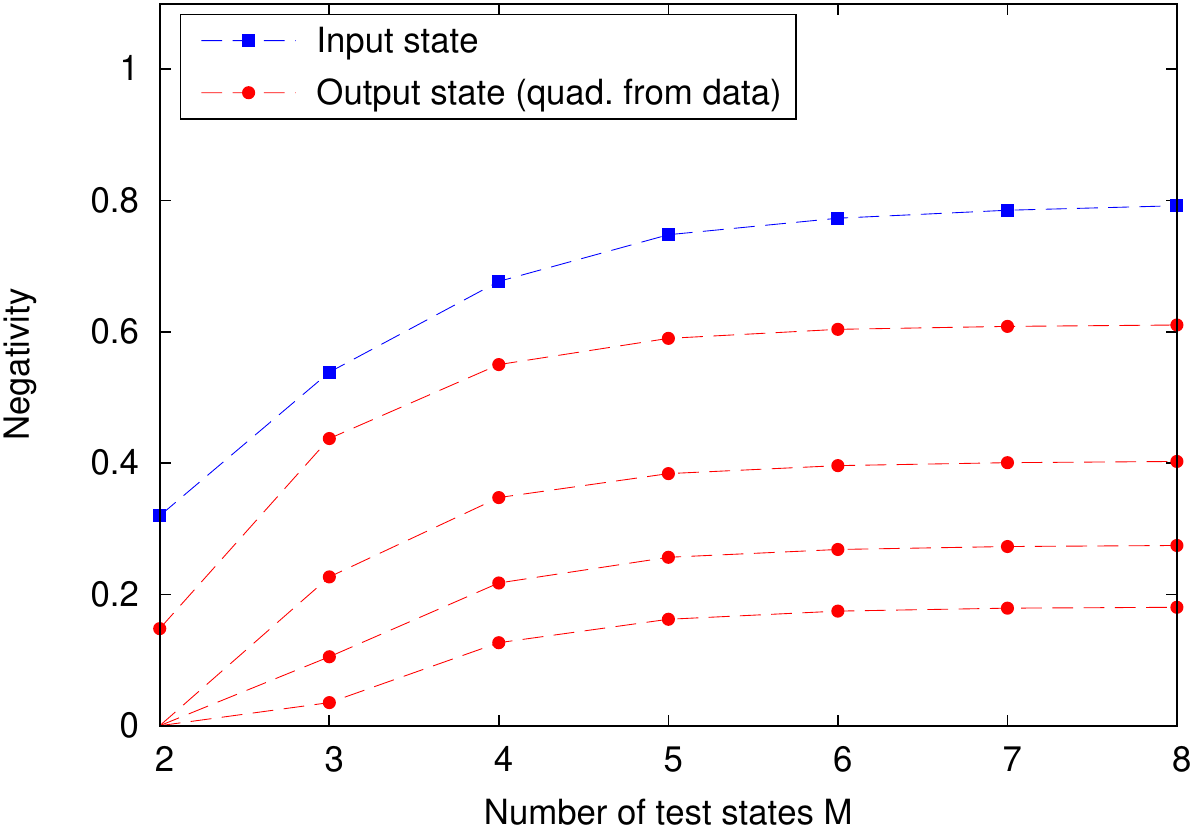}
	\caption[Benchmarking results using quadrature data directly]{Benchmarking results based on rotationally-symmetric test ensembles using direct quadrature measurement data. The top curve is the same as Fig. \ref{fig:negtomo}. The other curves are lower bounds to the negativity of $\rho_{AB}^\mathrm{out}$ based on (in descending order): (iii) quadrature values obtained directly from measurement data and (iv) quadrature data with error bars included in the optimization ($1\sigma$, $2\sigma$, and $3\sigma$ levels). For more than 3 test states, there is non-zero entanglement remaining even when including error estimates.}
	\label{fig:negquaddata}
\end{figure}

In cases (ii)-(iv), the choice of $N$ affects the strength of the optimization procedure itself \cite{killoran11a}, not just the density matrix. Truncating at a higher level would lead to better bounds on the entanglement. In all cases, we use the standard form results of Theorems \ref{thm:stform} and \ref{thm:tracenormstform}. Finally, although \cite{killoran11a} does not explicitly give a procedure for including error bars in the computation, it is straightforward to relax the framework and include them. We performed the optimization in Matlab, using the frontend YALMIP \cite{lofberg04a} and the solver SDPT3 \cite{toh99a}. The numerical precision of the optimization is estimated to be $\approx 10^{-3}$. The results of these computations for different sized rotationally-symmetric ensembles is shown in Figs. \ref{fig:negtomo} and \ref{fig:negquaddata}.

There are a number of things to point out about the results in Figs. \ref{fig:negtomo} and \ref{fig:negquaddata}. Primarily, we see that our benchmarking tools were successful in finding non-zero output entanglement in this phase-symmetric situation. Not only is the output state $\rho_{AB}^\mathrm{out}$ entangled for all $M$ when we use the tomographically reconstructed density matrix $\rho_0^\mathrm{out}$, but we can also see non-zero entanglement using the restricted set of measurements, including error bars. Furthermore, there is an evident advantage in using more than two test states, since it leads to higher entanglement in all cases. Indeed, it is not until $M>2$ that the $3\sigma$ level conclusively leads to non-zero entanglement in the output state. 

Another interesting result is that scenarios (i) and (ii) give very similar results for the entanglement when $M\geq 3$. This is likely due to the fact that the output state has low levels of excess noise. The quadratures would already reveal that such a state is close to a minimum uncertainty state. Additional measurement information therefore does not contribute significantly to our knowledge of the state. For other types of states, quadrature measurements and tomography should not be expected to yield similar results (though there may be other choices of measurement operators which work well). Note that we do not compare situations (i) and (iii) because of the extra phase shift included in the tomographic reconstruction outlined in Sec. \ref{ssec:dataacq}.

We also point out the advantage offered by phase-symmetric benchmarking and the standard form. Previously, entanglement quantification results were demonstrated for test ensembles consisting of only two or three states \cite{killoran10b, killoran11a}. Although the methods of \cite{killoran11a} apply to any number of (pure) test states, the computational resources required for more than three test states were prohibitive. Here, using the standard form, we have no problem pushing up to much higher numbers of states; Figs. \ref{fig:objfuncompare}-\ref{fig:negquaddata} contain results for up to 8 test states. In fact, with a desktop PC, we were able to calculate benchmarks for up to $10$ states, but these results are not included in the figures because they do not significantly change after $M=8$. Finally, we note that all the results for the output state entanglement are lower bounds. If we were to use a real entangled state instead of a virtual one, we would have access to more information, and the output entanglement would be higher. Even still, for benchmarking schemes based on a virtual entangled state, the entanglement quantification procedure yields useful and illuminating results.

\section{Conclusion}\label{sec:conclusion}

We have extended the methodology for entanglement-based quantum benchmarking to arbitrary finite ensembles of test states. This extension considers purifications of the test states, and we have outlined how to find strong choices for these purifications, leading to the best chance of certifying a device in the quantum domain. We also gave equations for reducing the test states to a standard form when certain phase symmetry conditions hold, allowing us to generate benchmarks from any single seed state. These theoretical tools were demonstrated using data from a real-world quantum memory and to quantitatively explore different benchmarking scenarios. Although we mainly considered test states which were of a continuous-variable nature, all of our results, except the standard form, apply to discrete states as well. Indeed, the same framework can be used to calculate fidelity-based benchmarks, which may be better suited to low-dimensional states \cite{killoran12b}. Since our benchmarking scheme works for arbitrary test states, it incorporates previous entanglement-based benchmarking tools \cite{rigas06a, rigas06b, haseler08a, haseler09a, haseler10a, killoran10b, killoran11a} into a unified framework. Together, these tools allow us to perform efficient benchmarking tests on quantum devices using realistic experimental resources. 

\acknowledgments{N.K. and N.L. acknowledge support from an NSERC Discovery Grant, Quantum Works, and Ontario Centres of Excellence. N.K. acknowledges additional support from the Ontario Graduate Scholarship Program. The ANU group acknowledges support from the Australian Research Council Centre of Excellence for Quantum Computation and Communication Technology (Project number CE110001027).}

\appendix
\section{Convexity proofs}\label{app:grammatconv}
In this appendix, we prove the following two statements about convexity: i) the objective function in Eq. (\ref{eq:puritydef}) is a convex function, and ii) the set of Gram matrices which are consistent with purifications of the test states is a convex set.
First, for positive semidefinite operators, the purity is equivalent to the squared Hilbert-Schmidt norm. By definition, all norms are convex, and taking the square preserves this convexity. Hence, our objective function is convex. 

To prove the second statement, let 
$\{\ket{\Gamma_k^0}_{A'A''}\}_{k=0}^{M-1}$ and $\{\ket{\Gamma_k^1}_{A'A''}\}_{k=0}^{M-1}$ be two sets of purifications of the test states $\{\rho_k\}_{k=0}^{M-1}$ (without loss of generality, we can consider the purifying system to be the same), with Gram matrices $G_0$ and $G_1$, respectively. Fix some $p\in[0,1]$ and take the convex combination of Gram matrices $G=pG_0+(1-p)G_1$. We need to show that there is some compatible set of purifications leading to the Gram matrix $G$. 

To do this, we define the following states:
\begin{align}
	\ket{\chi_k}_{A'A''A'''}:=\sqrt{p}\ket{\Gamma_k^0}_{A'A''}\ket{0}_{A'''}+\sqrt{1-p}\ket{\Gamma_k^1}_{A'A''}\ket{1}_{A'''},
\end{align}
where $\ket{0}_{A'''}$ and $\ket{1}_{A'''}$ are orthonormal states on some additional purifying system $A'''$. Tracing out all systems except $A'$, we find
\begin{align}
 \tr_{A''A'''}\ketbra{\chi_k}{\chi_k}_{A'A''A'''} = & ~p\tr_{A''}\ketbra{\Gamma_k^0}{\Gamma_k^0}_{A'A''}\nonumber\\
				         &~ +(1-p)\tr_{A''}\ketbra{\Gamma_k^1}{\Gamma_k^1}_{A'A''}\nonumber\\
				      = &~ \rho_k.
\end{align}
Thus, the $\ket{\chi_k}_{A'A''A'''}$ are purifications of the $\rho_k$. The elements of the corresponding Gram matrix are
\begin{align}
 \braket{\chi_k}{\chi_l} = &~ p \braket{\Gamma_k^0}{\Gamma_l^0} + (1-p) \braket{\Gamma_k^1}{\Gamma_l^1}\nonumber\\
			 = &~ p [G_0]_{kl} + (1-p) [G_1]_{kl},
\end{align}
which are exactly the elements of the convex combination $G$. Therefore, the set of Gram matrices which come from purifications of the test states is a convex set.

\section{Proof of standard form}\label{app:stformproof}
In this appendix, we give the proof of the standard form in Theorem \ref{thm:stform}. If Eq. (\ref{eq:taurotations}) holds, then $\tau_{AB}$ can be transformed by the unitary matrix
\begin{equation}\label{eq:Rdef}
	R = \bigoplus_{k=0}^{M-1}U_\theta^k
\end{equation}
to the form
\begin{align}\label{eq:Cdef}
	\mathcal{C}(\tau_{AB}) & = R^\dagger\tau_{AB}R\nonumber\\
	& = \frac{1}{M}\begin{bmatrix}
	                               	\tau_{00} & W_{01} & W_{02} & \cdots & W_{0,M-1}\\
					W_{0,M-1} & \tau_{00} & W_{01} & \cdots & W_{0,M-2}\\
					W_{0,M-2} & W_{0,M-1} & \tau_{00} & \cdots & W_{0,M-3}\\
					\vdots & \vdots & \vdots & \ddots & \vdots\\
					W_{01} & W_{02} & W_{03} & \cdots & \tau_{00}
	                               \end{bmatrix}
\end{align}
with $W_{ij}:=\tau_{ij}U_\theta^{j}$. The matrix $\mathcal{C}(\tau_{AB})$ has a \emph{block circulant} structure, meaning that each row of blocks is the same as the previous row, but shifted by one to the right. Now, let $\omega_M = \exp(i\theta)$ be the primitive $M$th root of unity and let 
\begin{equation}\label{eq:fourierdef}
    \left[F_M \right]_{ij} = \tfrac{1}{\sqrt{M}}\omega_M^{i\cdot j}
\end{equation}
be the (unitary) discrete Fourier transform matrix. Using a standard theorem on block circulant matrices \cite{davis94a}, $\mathcal{C}(\tau_{AB})$ can be block diagonalized by the matrix $F_M\otimes\iden_B$. Explicitly, 
\begin{align}\label{eq:Ddef}
	\mathcal{D}(\tau_{AB}) = & (F_M^\dagger\otimes\iden_B)\mathcal{C}(\tau_{AB})(F_M\otimes\iden_B)\nonumber\\
	 = & \bigoplus_{k=0}^{M-1}E_k,
\end{align}
with 
\begin{equation}\label{eq:Edecomp}
	E_k = \frac{1}{M}\sum_{l=0}^{M-1}\omega_M^{k\cdot l}W_{kl}
\end{equation}
Since the partial trace is unaffected by local unitaries on the same subsystem, we have
\begin{align}
	\sum_{k=0}^{M-1}E_k = &\tr_A\left(\sum_{k=0}^{M-1}\ketbra{k}{k}_A\otimes E_k\right) \nonumber\\
			    = &\tr_A\mathcal{D}(\tau_{AB})\nonumber\\
			    = &\tr_A\mathcal{C}(\tau_{AB}).
\end{align}
Comparing this with Eq. (\ref{eq:Cdef}), we conclude 
\begin{equation}
 \sum_{k=0}^{M-1}E_k = \tau_{00},
\end{equation}
which proves Eq. (\ref{eq:tau00decomp}). Finally, the positive semidefinite condition follows directly from the unitary equivalence of $\tau_{AB}$ and $\mathcal{D}(\tau_{AB})$.

\hfill$\square$


\section{Proof of trace norm in standard form}\label{app:tnormstformproof}
In this appendix, we prove Theorem \ref{thm:tracenormstform}, which gives a formula for the negativity involving the standard form. If $\tau_{AB}$ satisfies the symmetry condition in Eq. (\ref{eq:taurotations}), then so will the partial transpose $\tau_{AB}^{T_A}$. Therefore, $\tau_{AB}^{T_A}$ is unitarily equivalent to some block diagonal matrix $\mathcal{B}(\tau_{AB}^{T_A}) = \bigoplus_{k=0}^{M-1}\tilde{E}_k$. The trace norm is a unitarily invariant norm, so we must have
\begin{equation}
	||\tau_{AB}^{T_A}||_1 = \left|\left|\bigoplus_{k=0}^{M-1}\tilde{E}_k\right|\right|_1 = \sum_{k=0}^{M-1}||\tilde{E}_k||_1,
\end{equation}
which proves Eq. (\ref{eq:negstform1}). 

In order to determine the matrices $\tilde{E}_k$ in Eq. (\ref{eq:negstform2}), we first find an explicit expression relating the standard form matrices $E_k$ and the matrix $\tau_{AB}$. Define the matrix elements of a bipartite matrix $H$ by
\begin{equation}
	[H]_{ij,kl}:=\bra{i}_A\otimes\bra{j}_B H\ket{k}_A\otimes\ket{l}_B.
\end{equation}
For subsystem $B$, we work in the Fock basis, where the rotation operator $U_\theta$ is diagonal, with elements
\begin{equation}
	[U_\theta]_{jl} = \omega_M^{-j}\delta_{jl}.
\end{equation}
The unitary matrix $R$ in Eq. (\ref{eq:Rdef}), formed by taking powers of $U_\theta$, is also diagonal in this basis, with the following elements:
\begin{equation}
	[R]_{ij,kl} = \omega_M^{-i\cdot j}\delta_{ik}\delta_{jl}.
\end{equation}
Therefore, the Fock basis elements of the block circulant matrix $\mathcal{C}(\tau_{AB})$, defined in Eq. (\ref{eq:Cdef}), are given by
\begin{equation}
	[\mathcal{C}(\tau_{AB})]_{ij,kl} = \omega_M^{i\cdot j-k\cdot l}[\tau_{AB}]_{ij,kl}
\end{equation}

To get the standard form, we need to perform a Fourier transform on the block circulant matrix $\mathcal{C}(\tau_{AB})$ as in Eq. (\ref{eq:Ddef}). The elements of the discrete Fourier transform $F_M$ are given in Eq. (\ref{eq:fourierdef}). After substitution, we arrive at the matrix elements of the standard form:
\begin{align}
	[\mathcal{D}(\tau_{AB})]_{ij,kl} = & \frac{1}{M}\sum_{m=0}^{M-1}\sum_{n=0}^{M-1}\omega_M^{n\cdot k-i\cdot m}[\mathcal{C}(\tau_{AB})]_{mj,nl}\nonumber\\
	= & \frac{1}{M}\sum_{m=0}^{M-1}\sum_{n=0}^{M-1}\omega_M^{m(j-i)+n(k-l)}[\tau_{AB}]_{mj,nl}.
\end{align}
Since $\mathcal{D}(\tau_{AB})$ is block diagonal, $\mathcal{D}(\tau_{AB}) = \bigoplus_{k=0}^{M-1}E_k$, we have
\begin{align}\label{eq:Ekformula}
	[E_k]_{jl} := & [\mathcal{D}(\tau_{AB})]_{kj,kl}\nonumber\\
	 = & \frac{1}{M}\sum_{m=0}^{M-1}\sum_{n=0}^{M-1}\omega_M^{m(j-k)+n(k-l)}[\tau_{AB}]_{mj,nl}.
\end{align}
This is the final formula linking the elements of $\tau_{AB}$ and the matrices $\{E_k\}_{k=0}^{M-1}$. For completeness, the inverse formula is given by
\begin{equation}
	[\tau_{AB}]_{ij,kl}=\frac{1}{M}\sum_{m=0}^{M-1}\omega_M^{m(i-k)+k\cdot l-i\cdot j}[E_m]_{jl}.
\end{equation}

When $\tau_{AB}^{T_A}$ is used in this formula instead of $\tau_{AB}$, we find the relation
\begin{equation}
	[\tilde{E}_k]_{jl} := \frac{1}{M}\sum_{m=0}^{M-1}\sum_{n=0}^{M-1}\omega_M^{m(j-k)+n(k-l)}[\tau_{AB}]_{nj,ml}.
\end{equation}
Hence, the the matrix $\tilde{E}$ indexed by $(j+l-k)$ has elements
\begin{equation}
	[\tilde{E}_{j+l-k}]_{jl}=\frac{1}{M}\sum_{m=0}^{M-1}\sum_{n=0}^{M-1}\omega_M^{m(k-l)+n(j-k)}[\tau_{AB}]_{nj,ml},
\end{equation}
which is the same as Eq. (\ref{eq:Ekformula}), after interchanging the summation indices $m$ and $n$. Therefore, $[\tilde{E}_{j+l-k}]_{jl}=[E_k]_{jl}$ (in the Fock basis).


\begin{thebibliography}{46}
\bibitem[{foo({\natexlab{a}})}]{footnote1}
\bibinfo{note}{For example, when the entangled state in Eq.
  (\ref{eq:Mmixed_entstate}) has maximal entanglement, the purifications must
  be orthogonal, $\braket{\Gamma_k}{\Gamma_l}=\delta_{kl}$. But orthogonal
  purifications can be perfectly distinguished and reproduced by a particular
  MP channel, giving the same output as a perfect quantum memory. Thus, no
  benchmarking protocol where Eq. (\ref{eq:Mmixed_entstate}) is maximally
  entangled will be successful.}
\bibitem[{foo({\natexlab{b}})}]{footnote2}
\bibinfo{note}{Note that we assume the test states are not affected by an
  inactive memory.}
\end{thebibliography}
\end{document}